\documentclass[12pt]{iopart}
\usepackage{graphicx,amssymb}

\newcommand{\be}{\begin{equation}}
\newcommand{\ee}{\end{equation}}
\newcommand{\bea}{\begin{eqnarray}}
\newcommand{\eea}{\end{eqnarray}}

\renewcommand{\(}{\left (}
\renewcommand{\)}{\right )}
\renewcommand{\[}{\left [}
\renewcommand{\]}{\right ]}

\renewcommand{\Re}{\mathop{\rm Re}}
\renewcommand{\Im}{\mathop{\rm Im}}

\newcommand{\vp}{\varphi}
\newcommand{\Km}{\mathcal{K}}
\newcommand{\Wm}{\mathcal{W}}
\newcommand{\Vm}{\mathcal{V}}
\newcommand{\MM}{\mathcal{M}}

\begin{document}

\hfill arXiv:0801.2116
\\ \mbox{}
\hfill DESY 08-003 

\title{Successfully combining SUGRA hybrid inflation and moduli stabilisation}

\author{S~C~Davis$^1$  and M~Postma$^{2,3}$}

\address{${}^1$ Service de Physique Th\'eorique, 
Orme des Merisiers, CEA/Saclay, 91191 Gif-sur-Yvette Cedex, France}
\address{${}^2$ DESY, Notkestra\ss e 85, 22607 Hamburg, Germany}
\address{${}^3$ Nikhef, Kruislaan 409, 1098 SJ Amsterdam, The
Netherlands}

\eads{\mailto{sdavis@lorentz.leidenuniv.nl},
\mailto{postma@mail.desy.de}}

\begin{abstract}
Inflation and moduli stabilisation mechanisms work well independently, and
many string-motivated supergravity models have been proposed for them. However
a complete theory will contain both, and there will be (gravitational)
interactions between the two sectors. These give corrections to the inflaton
potential,  which generically ruin inflation. This holds true even for
fine-tuned moduli stabilisation schemes. Following a suggestion
by~\cite{anaXW}, we show that a viable combined model
can be obtained if it is the K\"ahler functions ($G= K+\ln |W|^2$) of the two
sectors that are added, rather than the superpotentials (as is usually done).
Interaction between the two sectors does still impose some restrictions on the
moduli stabilisation mechanism, which are derived. Significantly, we
find that the (post-inflation) moduli stabilisation scale no longer
needs to be above the inflationary energy scale.
\vskip 0.1in 

\noindent {\bf Keywords:} inflation, cosmology of theories beyond the SM
\end{abstract}

\section{Introduction}

Many attempts have been made to implement inflation in extensions of
the standard model, although to date there is still no model that is
truly convincing.  Supersymmetric (SUSY) theories appear to be more
promising. They include numerous moduli fields, i.e.\ scalar fields
which in the supersymmetric limit have an exactly flat potential, as
is required for slow-roll inflation.  Any one of these moduli fields
could play the role of the inflaton field.  As a concrete example we
will consider $F$-term hybrid inflation in this work. In the SUSY
limit it  has a flat direction, but when extended to include gravity
the situation is less rosy. The large energy density during inflation
breaks SUSY spontaneously, and supergravity (SUGRA) effects lift the
flatness of the moduli potential. This is the infamous
$\eta$-problem~\cite{copeland,dine}.

Furthermore, the particular form of the SUGRA potential means that all
other, non-inflationary sectors of the full theory will couple to the
inflation sector. The coupling will be small, in the models we
consider it is only of gravitational strength, but it can nevertheless
have large effects.  This is a generic problem for all inflation
models, and as we will see, this small coupling between the different
sectors frequently kills an otherwise good model. As a specific
example, we will study the effects of a modulus stabilisation sector
on 4D ${\mathcal N}=1$ SUGRA $F$-term hybrid inflation. All the other,
non-inflationary moduli fields must be fixed during inflation, and so
a full SUGRA theory must include additional physics to do this.  For
this we will consider  KKLT-like~\cite{KKLT} and KL-like~\cite{KL2}
moduli stabilisation schemes.  As we will see, the moduli sector gives
rise to additional --- and quite generically fatal --- corrections to
the inflaton potential. This raises the questions of whether the original
SUSY hybrid inflation model can actually be embedded in a full,
realistic theory, and if so, are its original predictions valid?  For
the answer to both these questions to be yes, the coupling between the
two sectors must somehow be minimal, so that neither the moduli
corrections to the inflation potential, nor the inflaton corrections to
the moduli stabilisation potential ruin the model. As we will show,
it is possible, but non-trivial, to achieve this.  

There are of course many other models of inflation, which offer
alternative approaches to the issue of moduli-inflation coupling. For
example, in modular inflation models the modulus field itself is the
inflaton~\cite{modular}. In a sense, the coupling is maximal ---
nevertheless successful (fine-tuned)  models have been
constructed~\cite{rt}.  In brane inflation models  the inflaton
potential arises from brane interactions, and depends explicitly on
the volume modulus.  Stabilising the modulus field then inevitably
gives a curvature correction to the inflation potential~\cite{KKLMMT}.
However explicit examples have been constructed where, for fine-tuned
parameters, the corrections to $\eta$ cancel to a high degree,
allowing inflation~\cite{baumann,krause}.  In contrast to the above models,
our strategy is to decouple the inflation and modulus sectors as much
as possible.  One advantage of this is that it also allows us to
decouple the scale of inflation from the gravitino mass scale.
At the cost of tuning, it is then possible to have the
gravitino in the phenomenologically favoured TeV range without the
need for low scale inflation.

The $\eta$-problem is a common feature of SUGRA inflation models.  To
illustrate it, consider a canonically normalised inflaton field with
$K= |\phi|^2$.  The inflationary potential is of the form
$V \sim \e^K V_* \sim V_*(1+ |\phi|^2 + \cdots)$, with $V_*$ the nearly
constant energy density driving inflation.  It follows that the
slow-roll parameter $\eta  = V''/V$ is of order unity, and slow-roll
inflation does not occur.  To avoid this conclusion one can fine-tune the
model such that the coefficient of the $|\phi|^2$-term in the
potential cancels.  More elegantly perhaps, one can try to achieve the
same using a symmetry.  An example of the latter approach is the
(accidental) Heisenberg symmetry of the K\"ahler potential in $D$-term
hybrid inflation~\cite{gaillard}.  In this paper we avoid the above
$\eta$-problem by using a shift symmetry for the inflaton, 
$\phi \to \phi + a$, which leaves the K\"ahler potential
invariant~\cite{modular,shift2}.  Since the inflaton field $\Re (\phi)$
no longer appears explicitly in the K\"ahler potential, the large mass
corrections to the inflaton field are avoided.

However, the shift symmetry does not kill all the corrections to the inflaton
potential.  In the presence of moduli fields $\eta$- (and $\epsilon$-)
problems appear again.  As a concrete example, consider the case of a single
modulus field $T$. If moduli fields are present, they need to be fixed during
inflation.   The modulus potential typically has a local minimum at finite
field value separated by a barrier from the global minimum at infinity.   The
classic example is the KKLT potential~\cite{KKLT}.  To assure that the modulus
does not run away during inflation the barrier should be large. This
is the case if the modulus mass is large $m_T > H_*$, with $H_*$ being
the Hubble constant during inflation~\cite{KL2}.  Now since the moduli
stabilisation mechanism breaks SUSY, there are soft corrections to the
inflaton potential, typically of $\Or(m_{3/2} H_*)$.  The flatness of
the inflaton potential is lost unless the gravitino mass is
sufficiently small $m_{3/2} < H_*$.  The problem with this is that 
one cannot tune the gravitino mass arbitrarily:  in a generic,
KKLT-like potential $m_{3/2} \sim m_T$, and a small gravitino mass is
at odds with keeping the modulus fixed. It is therefore difficult to
embed inflation in such a scheme.

A solution to the above moduli problem put forward by Kallosh and
Linde (henceforth denoted by KL)~\cite{KL}  is to fine-tune the
modulus potential so that $m_{3/2} \ll m_T$. Then if the Hubble
constant during inflation is between these two mass scales, the
modulus remains fixed while the soft  corrections to the inflaton mass
are small.  Such a set-up has the additional advantage that the
gravitino mass can be in the TeV range without the need for low scale
inflation.   KL gave an explicit realisation of this idea using a
racetrack potential for the modulus.  All problems then appear to be
solved, but this is deceiving.  Although the moduli corrections are
small after inflation thanks to the fine-tuning in the KL set-up, this
is not necessarily true during inflation.  During inflation the
modulus field $T$ is slightly displaced from its post-inflationary
minimum, disrupting the minute fine-tuning of the potential, with
potentially serious consequences.  Indeed, as we will show, in
$F$-term hybrid inflation the effects of the modulus displacement are
substantial, resulting in $\eta \approx -3$ and ruining inflation.
The need to include the dynamics of the modulus field during inflation
was previously noted in \cite{mfi,lalak}.

In this paper we will study $F$-term hybrid inflation, which serves to
illustrate all the observations made above. It is a multi-field model
of inflation, consisting of the inflaton field, and two oppositely
charged waterfall fields which are responsible for ending inflation.
When combined with a KKLT modulus sector, the corrections to both the
inflaton and the waterfall field potentials are large.  Although the
mass correction to the inflaton can be protected by a shift symmetry,
this is not the case for the waterfall fields, and as a result there
is generally no graceful exit from inflation.  Tuning the modulus
sector, as in the KL set-up, can reduce these corrections to a
harmless size.  However all of this is under the assumption that the
modulus $T$ is fixed during inflation.  Taking the modulus dynamics
into account we find that even in the fine-tuned KL-stabilisation
scheme the corrections are not harmless after all. On the contrary,
they prevent inflation from working.

In all previous studies of the effect of the moduli sector on
inflation~\cite{mfi,lalak,mdi,riotto}, the K\"ahler and superpotentials
of the modulus and inflaton sectors were simply added to get the
combined theory, i.e.\ take $W_{\rm total} = W_{\rm inf} + W_{\rm  mod}$
to get the full superpotential.  In this paper we instead 
multiply the superpotentials: $W_{\rm total} = W_{\rm inf} W_{\rm  mod}$, 
as proposed by Ach\'ucarro and Sousa~\cite{anaXW}.
As we will show, this greatly reduces the moduli corrections. Indeed
$F$-term hybrid inflation combined with KL, or even KKLT,  in this way
can give a viable inflation model.  Although multiplying superpotentials 
may sound odd at first, it is natural in a supergravity formulation in 
terms of the K\"ahler function $G = K +\ln |W|^2$.  Any supersymmetric
theory only depends on the K\"ahler- and  superpotential through the
combination $G$, suggesting that it is the only significant
quantity. Adding the K\"ahler functions of the two sectors is
equivalent to adding their K\"ahler potentials and multiplying their
superpotentials.

Adding K\"ahler functions has the nice property that a SUSY critical
point of the modulus sector is automatically a SUSY critical point of
the full theory as well~\cite{anaXW,binXW} --- this feature is at the
heart of the reduced moduli corrections.  In the limit of a small
gravitino mass, all the corrections to the inflaton potential are
small, including those due to the dynamics of the modulus field during
inflation.  The resulting inflationary model thus gives similar
inflationary predictions to the usual $F$-term hybrid inflation in the
absence of a modulus sector.  Although there are still some
constraints on the model parameters,  we want to stress that
successful inflation is achieved without the need for fine-tuning ---
this is in contrast to  most other combined inflaton-moduli models.
A notable feature of the model is that it is possible for the vacuum
modulus mass to be smaller than the Hubble scale during inflation,
without the modulus running off to infinity.

This paper is organised as follows.  In the next section we provide
the relevant background material.  We start with a short review  of
standard $F$-term hybrid inflation, both in a SUSY and SUGRA theories.
This is followed by a concise discussion of moduli stabilisation in
KKLT- and KL-style schemes.  In section~\ref{s:add} we discuss the
resulting model when the two sectors are combined by adding
superpotentials.  As we will see, even in the fine-tuned KL set-up
this does not give a working model.  In section~\ref{s:mult} we
combine the modulus and inflaton sectors by their multiplying
superpotentials, or equivalently by adding their K\"ahler-functions.
The modulus corrections to the inflaton potential now are under
control, and for a certain range of parameters we get successful
inflation. The parameter range for which the standard $F$-term hybrid
inflation predictions apply is determined in section~\ref{s:inf}.  We
end with some concluding remarks.

Throughout this article we will work in units with
$M_{\rm pl} = 1/\sqrt{8\pi G_N} =1$.

\section{Background} 

\subsection{SUSY $F$-term hybrid inflation}
\label{s:hybrid}

The superpotential for standard SUSY $F$-term hybrid inflation
is~\cite{hybrid1,hybrid2}
\be
W_{\rm inf} = \lambda \phi (\phi^+ \phi^- -v^2) \, .
\label{Winf}
\ee 
with $\phi$ the singlet inflaton field, and $\phi^\pm$ the waterfall fields
with charges $\pm 1$ under some $U(1)$ symmetry.  We can make
$\lambda$ real by an overall phase rotation of the superpotential,
whereas the phase of $v$ can be absorbed in the waterfall fields.
This is the convention we will use throughout this paper.  In
particular, in sections \ref{s:add} and \ref{s:mult} where we combine
inflation with a moduli stabilisation potential, all residual phases
reside in the moduli superpotential.  The scalar potential is
\be
V_{\rm inf}
= \lambda^2 |\phi|^2  \left(|\phi^+|^2 + |\phi^-|^2\right)
+ \lambda^2\left|\phi^+\phi^- - v^2\right|^2 + V_D \, . 
\ee
Vanishing of the $D$-term potential enforces $|\phi^+| = |\phi^-|$.  Inflation
takes place for $|\phi| > v$, during which the waterfall fields sit at the
origin $\phi^\pm=0$. The potential then reduces to a constant energy density
\be
V_{\rm inf} = V_* \equiv \lambda^2 v^4 \, ,
\label{VinfSUSY}
\ee
which drives inflation. The inflaton potential is flat at tree level, but
quantum corrections generate a slope for the inflaton field. The one-loop
potential is given by the Coleman-Weinberg formula~\cite{CW,loop}
\begin{equation}
V_\mathrm{loop} = \frac{1}{32 \pi^2} \mathrm{Str} M^2 \Lambda^2
+\frac{1}{64\pi^2} 
\mathrm{Str} M^4 \left(
\log \frac{M^2}{\Lambda^2} -\frac{3}{2}\right) 
\, ,
\label{CW}
\end{equation}
with the supertrace defined as 
$\mathrm{Str} f(M) = f(M_{\rm (boson)}) - f(M_{\rm (fermion)})$, and $\Lambda$
is the cut-off scale.
During inflation SUSY is broken and the masses of the waterfall field and
their superpartners are split
\be
m^2_\pm = \lambda^2 (|\phi|^2 \pm v^2) \, , \qquad
\tilde{m}^2_\pm = \lambda^2  |\phi|^2 \, ,
\label{masssplit}
\ee
giving a non-zero contribution to the logarithmic term in $V_{\rm loop}$.
Inflation ends when the inflaton drops below the critical value 
$|\phi| = v$, and one combination of the waterfall fields becomes tachyonic.
During the phase transition ending inflation the $U(1)$ symmetry gets broken
and cosmic strings form according to the Kibble
mechanism~\cite{kibble,strings}.

The predictions for the CMB power spectrum and spectral index are
\be
P = \frac{V}{150 \pi^2 \epsilon}
\, , \qquad
n_s = 1-\frac{d \ln P(N)}{dN}
\approx 1+2 \eta -6 \epsilon \, ,
\ee
evaluated at $N=N_* \sim 60$, where $N=-\log a$ is the number of $e$-folds
before the end of inflation. The slow-roll parameters are 
$\epsilon = (1/2)\({V'}/{V}\)^2$ and $\eta = {V''}/{V}$, with primes denoting
differentiation with respect to the canonically normalised real inflaton field
$\vp$, which for the above model is $\vp = \sqrt{2}|\phi|$.  The COBE
normalisation~\cite{COBE} for the power spectrum is 
$P \approx 4 \times 10^{-10}$, and WMAP3 results~\cite{WMAP3} give 
$n_s \approx 0.95 \pm 0.02$. We note however that if cosmic strings give a
minor contribution to the power spectrum, larger values of the spectral index
are favoured~\cite{urrestilla}.

We can get approximate analytical expressions in two limiting cases. For large
couplings $\lambda^2 \gtrsim 7.4 \times 10^{-6}$ inflation takes place for
large field values $\vp \gg v$, and the potential including loop corrections
approximates to
\be
V_{\rm inf} \approx V_* \left[ 1 + \frac{\lambda^2}{8 \pi^2}
 \log \frac{\lambda \vp}{\sqrt{2}\Lambda}\right] \, .
\label{Vlarge}
\ee
It follows that $N$ $e$-folds before the end of inflation, the inflaton field
is $\vp \approx \lambda \sqrt{N}/(2\pi)$. The prediction for the power
spectrum is  $P \approx 16 N_* v^4/75$, which when normalised to the COBE
scale gives  $v^2 \approx 5.6 \times 10^{-6}$. The spectral index is
$n_s \approx 1-1/N_* \approx 0.98$.  In the opposite limit, 
$\lambda^2 \lesssim 7.4 \times 10^{-6}$, inflation takes place for
inflaton values close to the critical value
$\vp_* \approx \vp_{\rm end} \approx \sqrt{2} v$.  Fitting the power
spectrum to the COBE normalisation now gives
$v^2 = 5.6 \times 10^{-6} [\lambda^2/(7.4 \times 10^{-6})]^{1/3}$, and
an approximately scale invariant spectrum $n_s \approx 1$.

Cosmic strings can contribute up to about 10\% (depending on the
angular scale) to the CMB power spectrum \cite{urrestilla,wyman,bjorn}.
This puts an upper bound on the string tension, and equivalently on
the symmetry breaking scale $v^2 < 10^{-5}$  --- $10^{-6}$, which
implies $\lambda < 10^{-3}$ --- $10^{-4}$~\cite{mairi,rachel}. However
there are ways to avoid cosmic string production, or at least relax
the bound~\cite{semilocal}.  In any case, the precise inflationary
predictions and the issue of cosmic strings is not the main point of
this paper.  Even if ruled out by future data, $F$-term hybrid
inflation still serves as a useful toy model to study the effects of a
moduli sector on inflation.  In particular it provides an explicit
example  for which multiplying superpotentials, instead of adding them,
helps to keep the moduli corrections under control.

\subsection{SUGRA $F$-term hybrid inflation}

Generically when an inflaton model is extended to include supergravity
corrections the potential develops a large curvature, resulting in a
slow-roll parameter $\eta \sim 1$ that is far too large for
inflation~\cite{copeland,dine}.  For $F$-term hybrid inflation with a
canonically normalised inflaton field this curvature correction
miraculously vanishes~\cite{hybridsugra}.  However, when higher order
corrections to the the K\"ahler potential are taken into account, or
when a modulus sector is included, this accidental cancellation is
destroyed, and the $\eta$-problem reappears.  It can be solved by
introducing a shift symmetry for the inflaton field into the
inflationary K\"ahler potential~\cite{modular,shift2}
\be
K_{\rm inf} = -\frac{(\phi-\bar \phi)^2}{2} + |\phi^+|^2 + |\phi^-|^2 \, .
\label{Kinf}
\ee
The canonically normalised inflaton, which is now 
$\varphi = \sqrt{2} \Re(\phi)$ (rather than $|\phi|$), does not appear
explicitly in the K\"ahler.

However, the SUGRA model with K\"ahler~\eref{Kinf} and
superpotential~\eref{Winf} still does not work.  The reason is that the mass
of the axion field $a = \sqrt{2} \Im(\phi)$ is tachyonic: $m_a^2 = -3
\lambda^2 v^4$. This problem is solved if we include an extra no-scale
modulus field $T$ in the model. Explicitly, take 
$K = -3 \ln (T+\bar T) + K_{\rm inf}$ and
\be
W_{\rm inf} =  \lambda_0 \phi (\phi^+ \phi^- -v_0^2) \, .
\label{Winf2}
\ee
The modulus field $T$ can arise in string theory as the breathing mode of
compactified extra dimensions; we will discuss it in more detail in
the next subsection. In the limit that $T$ is fixed we recover \eref{VinfSUSY} 
with $v=v_0$, and $\lambda = \lambda_0 (2\Re T)^{-3/2}$ the rescaled
coupling. The mass of the axion field is now positive definite 
$m_a^2 = 2 \lambda^2 v^4 (3+2\phi^2)$. The masses of the waterfall
fields are also altered
\be
m^2_\pm = \lambda^2 [\phi^2 + v^4 (1+\phi^2) \pm v^2(1+2\phi^2)] \, , \qquad
\tilde{m}^2_\pm = \lambda^2  |\phi|^2 .
\label{mpmSUGRA}
\ee
Since $v \ll 1$ the $v^4$ term is negligbly small. For $\lambda \lesssim 0.5$
we have $\phi^2 \lesssim 1$, and the other correction is also small. The
waterfield masses then reduce to the global SUSY results~\eref{masssplit}, and
the model approaches the SUSY limit. 

This is all very well, but in the above discussion we have neglected to
include a stabilisation mechanism for the modulus $T$. The full theory must
include additional potential terms, which break SUSY and are expected to give
corrections to the effective inflaton potential. This is
actually part of a wider issue, namely that inflation does not exist in
isolation --- it is part of a full theory containing other very high energy
physics (such a stabilisation mechanisms for moduli fields like $T$). Given
the restrictive form of SUGRA theories, interaction between different sectors
is unavoidable (gravity couples to everything). As we will see in
later sections, this can be catastrophic for many apparently good
theories, and leads to severe restrictions on others. Before discussing
the moduli corrections to inflation, we will first review moduli
stabilisation in the KKLT and KL set-ups.

\subsection{KKLT and KL moduli stabilisation}

KKLT devised an explicit method for constructing dS or Minkowksi vacua
in string theory \cite{KKLT}.  In their set-up all moduli fields are fixed by
fluxes \cite{gkp}, except for the volume modulus $T$ which is stabilised by the
superpotential
\be
\Wm_{\rm KKLT} = W_0 + A \e^{-a T} \, , \qquad \Km = - 3 \log[T + \bar T] \, ,
\label{WKKLT}
\ee 
where $W_0$ comes from fluxes, and the non-perturbative exponential term from
gaugino condensation or alternatively from instanton effects. For a general
SUGRA theory, the $F$-term potential is
\be
\Vm_F = \e^{\Km} \left( \Km^{I \bar J} 
D_I \Wm \bar D_{\bar J} \bar \Wm - 3|\Wm|^2\right)
\ee
with $D_I \Wm= \Wm_{\!, I} + \Km_I \Wm$.  The minimum of the above
superpotential~\eref{WKKLT} is SUSY preserving and AdS.  However, we
require a Minkowski or dS vacuum with a small cosmological constant to
desribe our universe.  This can be obtained by adding an uplifting
term, which then gives a minumum in which SUSY is broken. In the
original KKLT paper an anti-D-brane was used for
uplifting. Alternatively a $D$-term can be used~\cite{bqk} although
additional meson fields are required to implement
this~\cite{ana}. $D$-term uplifting has the advantage that the full
theory can still be described by SUGRA, whereas the KKLT uplifting
term breaks SUSY explicitly.  In this paper we assume any lifting term
takes the form
\be
\Vm_{\rm lift} \propto \frac{K_T^2}{\Re f(T)} \, ,
\label{VLift}
\ee
where $f(T) \propto T$, or is a constant. This gives the correct form for the
KKLT lifting $\Vm_{\rm lift} \propto (\Re T)^{-n}$ with $n=2,3$. The
$D$-term will also include the meson fields, although $\Vm_{\rm lift}$
is qualatively the same (at least for the analysis of this paper). 

Alternatively one can introduce an uplifting $F$-term sector, such as an
O'Raifeartaigh~\cite{O'Raifeartaigh} or ISS~\cite{ISS} sector. An
explicit example of this is the O'KKLT model~\cite{KL}, in which a
minimal O'Raifeartaigh sector is added to~\eref{WKKLT}. In this paper
we will implement the theory  with a no-scale K\"ahler. The full
moduli stabilisation sector is then
\be
\Km = -3 \ln \left[T + \bar T - \frac{K_{O'}}{3}\right]
 \, , \qquad 
\Wm = \Wm_{\rm KKLT}+ \Wm_{O'}
\label{OKL}
\ee
with
\be 
\Km_{O'} = S \bar S - \frac{(S \bar S)^2}{\Lambda_{s}^2} 
\, , \qquad \Wm_{O'} = -\mu^2 S \, .
\label{O'}
\ee
The O'Raifeartaigh sector breaks SUSY and lifts the AdS vacuum to
Minkowski. There is then no need for a separate non-$F$ lifting term in the
theory. 

The resulting stabilisation potential $\Vm_{\rm mod} = \Vm_F +\Vm_{\rm lift}$
has only one scale $m_T \sim m_{3/2}$. The Minkowski
minimum is separated  from $T=\infty$ by a barrier of height 
$\Vm_{\rm max} \sim m_T^2$. The barrier needs to be higher than the
inflationary scale, otherwise the moduli will roll off to infinity and
the internal space will be decompactified, which gives the bound 
$H_* < m_{3/2}$ on the inflationary scale \cite{KL2}. 

KL devised a moduli stabilisation scheme that circumvents the above
bound on the Hubble scale during inflation~\cite{KL}.  Instead of the
KKLT superpotential they use a modified racetrack superpotential
\be \Wm_{\rm KL} =W_0+ A \e^{-a T} +B \e^{-b T} \, .
\label{WKL}
\ee 
The extra parameters in the superpotential allow us to tune
$\Wm_{\! ,T} = \Wm=0$, giving a metastable SUSY Minkowski vacuum without the
need for a lifting term. As it stands, the model has $m_{3/2}=0$.  This can be
avoided by slightly perturbing the Minkowski solution to obtain an AdS minimum
$V \sim -m_{3/2}^2 \ll m_T^2$, which is then uplifted to a SUSY breaking
Minkowski vacuum.  Uplifting can be done with a small KKLT lifting term, or
alternatively by adding an uplifting $F$-term sector~\eref{O'}, as was used in
section 3  of~\cite{KL}. If the SUSY-breaking scale is small, we have 
$T \Wm_{\! ,T} \sim \Wm \sim m_{3/2} T^{3/2}$ and the gravitino mass is far
smaller than the modulus mass scale, which is typically set by $W_0$ in the
superpotential. It is then possible to have 
$m_{3/2}^2 \ll H_*^2 \ll \Vm_{\rm max} \sim m_T^2$, which opens the
possibility of having inflation with fixed moduli but small soft corrections
to the inflaton potential. Note that such a scenario cannot be implemented
with an uplifting $D$-term. In this case gauge symmetry implies that the
Minkowski solution  $\Wm_{\! ,T} = \Wm=0$ is obtained along a flat direction
in the meson-modulus field space.  As a result, after perturbing the solution
and uplifting to Minkowski, only one modulus mass eigenstate is
large. The other is only $\Or(m_{3/2})$, and so the barrier height
along the previously flat direction is also small 
$\Vm_{\rm max} \sim m_{3/2}^2$, even when the modulus mass is large 
$m_T \gg m_{3/2}$.

The above model \eref{OKL} uses a slightly different $\Km$
to~\cite{KL}, although it has similar properties. We have chosen the
above K\"ahler to simplify the analytical expressions. But we want to
emphasise that the exact way the modulus potential and the
O'Raifeartaigh section are combined does not significantly affect
inflation.  For that matter, the uplifting sector does not have to be
O'Raifeartaigh either, but can be some other $F$-term SUSY breaking
sector such as the ISS model.  The differences in the resulting
potential will be of order $\Or(m_{3/2}^2)$,  and as long as
$m_{3/2} \ll H_*$ such differences are irrelevant during
inflation.  As we will see in section~\ref{s:mult}, whether the
uplifting is $F$-term or not can make a major difference.  For the
case where the modulus and inflaton sector are combined by adding
their respective  K\"ahler functions it is the difference between a
viable model and no model at all.

\section{Combining inflation and moduli stabilisation by addition}
\label{s:add}

The usual way to combine the models of the previous sections is
to add the respective superpotentials $W = \Wm + W_{\rm inf}$. Here
$\Wm$ is the modulus superpotential, either KKLT~\eref{WKKLT} or
KL~\eref{WKL}, possibily including an $F$-term O'Raifeartaigh lifting sector. 
For the K\"ahler potential we consider the simplest possibility
\be K = -3 \ln \left[ X \right] + K_{\rm inf} \, ,
\label{Kadd}
\ee
with
\be
X = T+ \bar T - \frac{\Km_{O'}}{3} \, .
\ee
If uplifting is achieved via an anti-D-brane or $D$-term, $\Wm_{O'}$ and
$\Km_{O'}$ are simply set to zero. To verify that the qualitative results are
independent of the exact form of the K\"ahler, we also consider the more
general expression
\be
K = -3 \ln \left[X - X^{\alpha} \frac{K_{\rm inf}}{3}\right] \, .
\label{Ka}
\ee
For $\alpha =0$ this gives a fully no-scale K\"ahler potential: 
$K_a K^{a \bar b} K_{\bar b} =3$ with $a,b$ running over both moduli and
inflaton fields.

Slow-roll inflation with a scale invariant spectrum of perturbations
requires $\epsilon,\eta \ll 1$.  Hence we have to make sure the moduli
induced corrections to the slope and curvature of the inflaton
potential are sufficiently small.  The corrections to the masses of
the waterfall and axion fields must also be small.  If the mass
corrections to the waterfall fields are too large and positive
definite, they prevent $\phi^\pm$ becoming tachyonic, and there is no
exit from inflation.  Alternatively, if the corrections are large and
tachyonic the system ends up in the wrong vacuum.  Furthermore, the
axion mass has to be positive definite during inflation, which is not
automatic.  For the moment we work in the approximation that the
moduli are fixed at the minimum $T=T_0$  during inflation.  At the end
of this section we will drop this assumption, and analyse its
implications.

For either choice of K\"ahler we find there are corrections to the
slope of the inflationary potential~\cite{mfi}. For \eref{Kadd}, the full
$F$-term potential for the combined theory is 
\be
V_F = e^{K_{\rm inf}} \Vm_{F} + \Vm_{\rm lift}
+ e^{K}|\partial_i W_{\rm inf}+ K_i (W_{\rm inf} + \Wm)|^2
+V_{\rm mix} \, ,
\ee
which is roughly the sum of the potential for the inflation and moduli sectors
(with some rescaling), and the additional mixing terms
\be
\fl
V_{\rm mix} = 2 e^{K} \Re 
[(\Km^{I\bar J}D_I\Wm \Km_{\bar J} -3 \Wm) \bar{W}_{\rm inf}) ]
+ e^{K} (\Km^{I\bar J} \Km_I \Km_{\bar J}-3) |W_{\rm inf}|^2 \, .
\label{Vadd}
\ee
The index $i$ runs over the inflation sector fields, while $I,J$ run over the
moduli sector fields. During inflation all $K_i=0$, and the SUGRA $K_i W$
corrections to SUSY inflation vanish. Furthermore, for a no-scale
moduli K\"ahler~\eref{Kadd} the second term of $V_{\rm mix}$ is identically
zero. The K\"ahler potential~\eref{Ka} gives rise to similar mixing terms.

Much of the moduli interaction effectively re-scales the inflationary
parameters, and so it is convenient to introduce
\be
\fl
\lambda = \frac{\lambda_0}{X^{3\alpha/2}} \, , \qquad 
v^2 = \frac{v_0^2}{X^{1-\alpha}}  \, , \qquad 
V_* = 3 H_*^2 = \lambda^2 v^4 \, , \qquad
\vp = \sqrt{2X^{\alpha-1}} \Re \phi
\, .
\ee
These apply to the general K\"ahler~\eref{Ka}, and also to~\eref{Kadd} if 
$\alpha$ is set to 1. In both cases the inflationary potential reduces to
\be 
V_{\rm inf} = V_* + \Vm_{\rm mod}
+ \frac{\sqrt{2}\Re (\Wm_{\! ,T})}{\sqrt{X}} \lambda v^2 \vp
\label{VinfA}
\ee
with $\vp$ the canonically normalised inflaton. The inflaton independent
modulus potential is $\Vm_{\rm mod}(T) = \Vm_F +\Vm_{\rm lift}$. We see that a
nearly  flat inflaton potential, with $\epsilon \ll 1$, requires
$V_{\rm mix} \propto \Re \Wm_{\! ,T}$ to be small. This can be achieved either
be making $|\Wm_{\!  ,T}|$ small (which is the case for the two-scale KL-style
stabilisation), or by having $\Wm_{\! ,T}$ imaginary, i.e.\ having a phase
difference between the inflation and moduli superpotentials.

We also need to check that the corrections to the masses of the waterfall
fields do not radically change the ending of inflation, and that the axion
$a=\sqrt{2} \Im \phi$ remains stable. We introduce the mass scales
\be
m = \frac{\Wm}{X^{3/2}} \, , \qquad
m' = \frac{\Wm_{\!, T}}{\sqrt{X}}  \, , \qquad
\MM = \frac{\sqrt{X} \Wm_{\! , TT}}{3} \, .
\ee
Up to small $\Or( \e^{K_{\rm inf}})$ corrections $|m|  \approx m_{3/2}$ is the
gravitino mass after inflation, and in a KL-style scheme $|\MM|
\approx m_T$ the modulus mass.  For KKLT we still have $|\MM| \sim
m_T$. For the K\"ahler \eref{Kadd} with canonically normalised
inflaton sector fields the masses of the axion and waterfall fields are
\be
\fl
m_a^2 = 2\lambda^2 v^4 (3+2\phi^2) + 2\Vm_{F} +4|m|^2  
-4 \Re[2m-m'] \lambda v^2 \phi \, ,
\label{ma}
\ee
\be
\fl
m_\pm^2 = \lambda^2\phi^2 
\pm \lambda^2 v^2 \left|1  - \frac{2m -m'}{\lambda v^2}\phi+2 \phi^2 \right|
+ \Vm_{F} +|m|^2 + \lambda^2 v^4 (1+\phi^2) 
+ 2\lambda v^2 \Re [m'-m] \phi \, .
\label{mpa}
\ee
For a one-scale KKLT-like moduli sector $m, {m'} \sim m_{3/2} \sim m_T$.  The
requirement that the moduli remain fixed during inflation, i.e.\ 
$H_*^2 <  \Vm_{\rm max} \sim  m_T^2$, implies that the 
$\Or(m,m')$  moduli sector corrections to $m_\pm^2$
dominate, preventing a gracefull exit from inflation. 
A further problem for models which use a D-brane or $D$-term lifting term 
$\Vm_{\rm lift}$ is that the axion and waterfall masses recieve large
tachyonic contributions from the moduli sector $F$-term potential 
$\propto \Vm_F \sim -3m_{3/2}^2$. For $F$-term lifting $\Vm_F=0$ in
the Minkowski vacuum after inflation, and so the contribution of
$\Vm_F$ during inflation is small.

In principle, all these problems can be avoided with sufficient fine-tuning,
although the single mass scale superpotential~\eref{WKKLT} does not contain
enough parameters. Hence we must switch to a two-scale KL moduli stabilisation
scheme, which is tuned so that $\Wm=\Wm_{\! ,T} \approx 0$ and thus 
$m , m', \Vm_F \approx 0$. The moduli corrections to the waterfall~\eref{mpa}
and axion~\eref{ma} field masses, as well as to the inflaton
potential~\eref{VinfA}, are then negligibly small during inflation.

So it appears that the potential can be kept flat and the mass corrections
small in a KL-style set-up.  But as we will now show this is not the final
picture. In the above analysis we assumed $T$ was fixed at the minimum of
$\Vm_{\rm mod}$. However, no field is truly fixed at a constant value during
inflation, and in particular the modulus minimum will shift slightly during
inflation.  Taking the dynamics of the modulus into account, we will now show
that it produces siginificant curvature corrections to the potential, and
consequently gives too large a value for $\eta$~\cite{mfi}.  To do so we Taylor
expand the potential \eref{VinfA} in $\delta T=T -T_0$, with as before $T_0$
the modulus value that minimises the {\em post}-inflationary potential:
\be
\fl
V_{\rm inf} = V_*(T_0) + \Vm_{\rm mod}(T_0) 
+  \frac{2\Re \Wm_{\! ,T}(T_0)}{X^2} \lambda_0 v_0^2  \phi
+\delta V_{\rm inf}
+ \Or \! \left(|\delta T|^3, \lambda_0 v_0^2 \phi |\delta T|, 
V_* |\delta T|\right)
\label{dT1}
\ee
where 
\bea
\delta V_{\rm inf} &=&
\Vm_{{\rm mod},T \bar T} \, \delta T \overline{\delta T}
+\Re [\Vm_{{\rm mod},TT} \, \delta T^2]
\nonumber \\ && {}
+2\left[ X \Re (\Wm_{\! ,TT} \, \delta T)  -
4 \Re (\Wm_{\! ,T}) \Re (\delta T) 
\right]\frac{ \lambda_0 v_0^2}{X^3}  \phi
\eea
gives the leading order corrections to $V_{\rm inf}$ from the variation of
$T$. Now for KL  $|\MM| \gg |m|, |m'|$, hence this reduces to
\be
\delta V_{\rm inf}
\approx 3\frac{|\MM|^2}{X^2} |\delta T|^2 
+ \frac{3 \sqrt{2} \lambda v^2 \vp}{X}\Re[\MM \, \delta T] \, .
\ee
Minimising with respect to $\delta T$ we find
\be
\frac{\delta T}{X} \approx - \frac{\lambda v^2 \vp}{\sqrt{2} \MM}
\ee
which is small (as expected). However when this is substituted back
into the above potential, it produces a large negative
inflaton mass
\be
\delta V_{\rm inf} \approx -\frac{3}{2} V_* \vp^2 \, .
\label{dT2}
\ee
The $\eta$-problem rears its head again:
$\eta = V_{\! , \vp \vp}/V \approx -3$.  For KL without the SUSY
breaking O'Raifeartaigh sector the above expressions are exact, while
an uplifting sector --- O'Raifeartaigh or otherwise  --- gives rise to
small $\Or(m_{3/2}^2)$ corrections (both due to the above $\delta T$
expression, as well as the displacement of e.g.\ the
O'Raifeartaigh field $\delta S$). The large slow-roll parameter rules
out $F$-term hybrid inflation with KL moduli stabilisation. The reason
for the large corrections, even in the fine-tuned KL set-up is that
although  $\Wm \sim \Wm_{\! ,T} \approx 0$ are small, 
$|\Wm_{\! ,TT}|^2 = 3X \Vm_{{\rm mod}, T \bar T} + \Or(\MM m_{3/2})$
is not.  In the Minkowski vacuum after inflation the potential is
fine-tuned so that $m_{3/2}^2 \ll m_T^2$,  but during inflation, due
to the small displacement of the modulus field, this tuning is
disrupted, and corrections are large.

For the more general K\"ahler~\eref{Ka} the inflaton potential is still given
by~\eref{VinfA}. The waterfall masses take the form
\bea \fl
m^2_\pm &=&
\frac{\lambda^2 \vp^2}{2}
\pm \lambda^2 v^2
\left|1 +  
\frac{(1+2\alpha)m' - 6\alpha m}{3 \sqrt{2} \lambda v^2} \vp
+\alpha \vp^2 \right|
+ \frac{2+\alpha}{3} \Vm_{F} + \frac{2(1-\alpha)}{3} \Vm_{\rm lift} 
\nonumber \\ \fl && {}
+ \alpha |m|^2 
+ \frac{\sqrt{2} \lambda v^2 }{3} \Re[(2+\alpha)m' -3 \alpha m] \vp
+ \lambda^2 v^4 \left(\frac{2}{3} + \frac{\alpha \vp^2}{2}\right)
  \, .
\eea
In general, the model will have all the same problems as that arising from
the simpler K\"ahler~\eref{Kadd}, and one-scale KKLT-style moduli
stabilisation superpotentials are ruled out. It is interesting to note
that for a no-scale $\alpha=0$ model most of the corrections to $m_\pm$ cancel
(compare with the $D$-term inflation model proposed in~\cite{mdi}). In
particular, all the $m$, $\Vm_F$ and $\Vm_{\rm lift}$ corrections
disappear. It would seem that we then only need to impose the single
fine-tuning $m' \approx 0$, to obtain a viable inflation
model. Unfortunately the KKLT superpotential~\eref{WKKLT} does not
have enough freedom to do this, and viable inflation is not
obtained. Furthermore, the above discussion does not take into account
the varation of $T$ during inflation. The above analysis of $\delta T$
also applies for the more general K\"ahler~\eref{Ka}, and so it too is
ruled out.

To conclude, $F$-term hybrid inflation does not work for either KKLT- or
KL-style moduli stabilisation, no matter what the form the K\"ahler takes, at
least if we combine the inflation and modulus sector by adding
superpotentials. In fact, if more exponential terms are added to the moduli
stabilisation superpotential, its first three derivatives are appropriately
tuned, and the K\"ahler is carefully choosen, the moduli dynamics could be
different to those used to get~\eref{dT2}. A viable model of inflation could
concievably be constructed, although it is hard to justify all the
fine-tuning. Furthermore, there is no guarantee that additional problems will
not arise as a result of this tuning. We will not consider such as set-up
here, and will instead turn to a much more elegant solution.

\section{Combining inflation and moduli stabilisation by multiplication}
\label{s:mult}

The inflaton and modulus sectors can also be combined by multiplying their
superpotentials.  Although due to its unfamiliarity this seems strange at
first, we argue that from a supergravity point of view it is a rather natural
thing to do.  Multiplying superpotentials greatly reduces the mixing between
sectors~\cite{anaXW,binXW}.  Indeed, as we will discuss in this section
$F$-term hybrid inflation combined in this way with KL or even a 
KKLT moduli sector gives a viable inflation model.

The supergravity formulation in terms of $K$ and $W$ is redundant, as a
K\"ahler transformation leaves the theory invariant. Instead the theory can
be formulated in terms of single K\"ahler invariant function 
$G = K +\ln|W|^2$, which is known as the K\"ahler function.   The kinetic
terms and $F$-term potential are then given in terms of $G$ only.  This
suggests that the K\"ahler function is a more ``fundamental'' or ``natural''
quantity to consider.   Hence when combining sectors, it may be argued that
one should add their respective K\"ahler functions, which corresponds to
adding K\"ahler potentials and multiplying superpotentials.

For the combined theory we then take $G = G_{\rm mod} + G_{\rm inf}$. The
reduced inflaton-moduli interactions are a result of the following property.
Consider a SUSY critical point $T=T_0$ of the modulus sector $\partial_T
G_{\rm mod} (T_0) =0$, which corresponds to a SUSY extremum of the moduli
potential.  It can easily be shown that this is then a SUSY critical point of
the full theory as well  $\partial_T G (T_0) =0$~\cite{anaXW,binXW}.
This is exactly what we want, as it implies that the modulus minimum is not
shifted during inflation. The $\delta T$ corrections to the potential, which
were fatal when adding superpotentials, are then absent. Of course, with SUSY
broken in the modulus sector the minimum of the modulus potential is not
exactly in a critical point.  But in the KL-like set-up the deviations away
from the SUSY critical point are small, of the order of the small gravitino
mass.  Consequently we expect the modulus field to be nearly constant during
inflation, and the corresponding correction to the potential to be suppressed
by the smallness of the gravitino mass. As we will see, this is indeed
the case.

One disadvantage of the K\"ahler function formulation of SUGRA is that it
is ill defined whenever $W=0$.  This presents a problem for $F$-term
hybrid inflation, as the inflationary superpotential \eref{Winf} is zero after
inflation.  To solve this problem we ``correct'' the superpotential by adding
a constant
\be
W_{\rm inf} = \lambda_0 \phi (\phi^+ \phi^- -v_0^2) - C \, .
\label{Winfx}
\ee 
Here we will assume that $C$ is real and positive, although generalisation of
the analysis to include a phase is straightforward. The constant $C$
is of course irrelevant in the IR global SUSY limit, whereas in the UV
regime it makes the model well behaved.  Similarly, for the modulus
potential we cannot take the supersymmetric KL limit, a finite amount
of SUSY breaking (explicitly provided in \eref{WKL} by an
O'Raifeartaigh sector) is required. The effective superpotential of
the model with the modulus included is now
\be
W = \Wm \, W_{\rm inf} \, .
\label{Wmult}
\ee
For the K\"ahler potential we still use \eref{Kadd} with canonically
normalised inflaton fields.  To test the dependence of the results on the
exact form of the K\"ahler we also give the results for the general expression
\eref{Ka}.

For the minimal K\"ahler \eref{Kadd} the potential that follows from
\eref{Winfx},\eref{Wmult} is
\be
V = e^{K_{\rm inf}} |W_{\rm inf}|^2 \Vm_F 
 + e^{\Km} |\Wm|^2  e^{K_{\rm inf}} 
\left|\partial_i W_{\rm inf}+ K_i W_{\rm inf} \right|^2
+\Vm_{\rm lift} \, .
\ee
As advertised, the mixing between the inflaton and modulus sector is
drastically reduced compared to the case of adding superpotentials
\eref{Vadd}.   The main effect is just a re-scaling of the potential.
We define the re-scaled quantities
\be
\fl
\lambda = \frac{\lambda_0 |\Wm|}{X^{3\alpha/2}} \, , \qquad 
v^2 = \frac{v_0^2}{X^{1-\alpha}}  \, , \qquad 
\Vm_{\rm mod} = C^2 \Vm_F + \Vm_{\rm lift}\, , \qquad 
\vp = \sqrt{2 X^{\alpha-1}} \Re \phi \, .
\label{par1}
\ee
$V_* = 3H_*^2 = \lambda^2 v^4$ is then the rescaled inflationary potential
driving inflation, while $\Vm_{\rm mod}$ is the full rescaled modulus
stabilsation potential after inflation.  The field $\vp$ is the real, 
canonically normalised, inflaton field.  As before, the expressions for
\eref{Kadd} correspond to $\alpha = 1$. We also define the mass scales
\be
m = \frac{C|\Wm|}{X^{3/2}} \, , \qquad 
\MM = \frac{C\sqrt{X} |\Wm_{\! , TT}|}{3} \, ,
\label{par2}
\ee
which can respectively be thought of as the gravitino and moduli mass in
the vacuum, after inflation. With these definitions the potential during
inflation for both \eref{Kadd} and \eref{Ka} reduces to
\be
V_{\rm inf} = V_* + V_{\rm mod}
 \left(1 + \frac{\lambda v^2 \vp}{\sqrt{2} m}\right)^2
- \frac{\lambda v^2 \vp}{\sqrt{2} m} 
\left(2 + \frac{\lambda v^2 \vp}{\sqrt{2} m}\right) 
\Vm_{\rm lift} \, .
\label{VinfM}
\ee
We see that if a seperate lifting term is present (either an anti-D-brane or a
$D$-term), its potential $\Vm_{\rm lift} \sim  m_{3/2}^2$ gives a large
negative contribution to $\eta$.  This holds for both the KKLT and KL
superpotential, and so all our moduli stabilisation scenarios with non-$F$
lifting terms are incompatible with $F$-term hybrid inflation. In the
remainder of this section will thus focus on the case of $F$-term lifting
with $\Vm_{\rm lift} =0$.

In the limit that the modulus remains fixed during inflation $\Vm_{\rm mod}=0$
for $F$-term lifting, and there are no corrections to the inflaton potential
at all. This is in sharp contrast to the potential obtained when adding
superpotentials~\eref{VinfA}. Although the modulus is not truly fixed during
inflation, we will see below that the corrections to this assumption are small.

In multiplying the superpotentials, our intention was to reduce the
effect of the moduli sector on inflation. We see from~\eref{VinfM} that a
beneficial side effect of this is that the inflaton enhances the moduli
stabilisation. In particular the barrier height for the moduli stabilisation
potential is now 
\be
V_{\rm max} \sim \MM^2\left(1 + \frac{\sqrt{3} H_* \vp}{\sqrt{2} m}\right)^2
\, .
\label{enhanced}
\ee
Hence we expect the moduli to remain near their minimum during inflation if 
$\MM \gg H_*$ (as is usually assumed), or if $(\MM/m) \vp \gg 1$. Since
$\vp > \vp_{\rm end} \sim v$, the moduli should be stable thoughout
inflation if either
\be
(a) \quad \MM \gg H_*  \qquad \mbox{or} \qquad 
(b) \quad \MM \gg \frac{m}{v} \gtrsim 4 \times 10^{2} \, m \, .
\label{bound5}
\ee
Significantly, the second possibility does not depend on the Hubble constant
during inflation, and so having $H_* > \MM$ is not a
problem. The $H_* < \MM$ bound was a major motivation for the KL scenario,
and its removal suggests that a two-scale, KL-style moduli sector is
no longer needed. However, while the bound (\ref{bound5}b) is easily
satisfied for KL, it cannot be satisfied by KKLT. Hence it seems that
a two-scale KL-like moduli sector is needed after all, although not
necessarily  for the reasons that were originally envisaged.

For the simplest K\"ahler \eref{Kadd} the waterfall field masses are
\be
m_\pm^2 = \lambda^2 \phi^2 
\pm  \lambda^2 v^2 \left(1 +\frac{2m}{\lambda v^2}\phi + 2\phi^2 \right)
+(m + \lambda v^2 \phi)^2 + \lambda^2 v^4 \, .
\label{mpm1}
\ee
In the limit
\be
m \approx m_{3/2}  \ll \frac{\lambda v^2}{\vp}
\label{bound}
\ee
the moduli corrections are subdominant, and inflation ends as in usual
hybrid inflation. From the COBE normalisation it follows that $v^2 \ll 1$ 
and all $v^2$ corrections can be neglected as well.  For a
KKLT-style superpotential~\eref{WKKLT} with $m \sim \MM$, it is
difficult to satisfy both of the above bounds~\eref{bound5},
\eref{bound} simultaneously, and most vlaues of $\MM$ are ruled
out. For smaller values of $\lambda$ (for which
$\vp_* \ll 1$) there is a small window of parameter space 
$H_* \ll \MM \ll H_*/\vp_*$ where inflation will be viable.  For a two-scale
KL-style scenario there is more room to satisfy the bounds
\eref{bound5}, \eref{bound}, but at the cost of fine-tuning the potential.

For the more general K\"ahler \eref{Ka} the waterfall masses are instead
\bea
m^2_\pm &=&
\frac{\lambda^2 \vp^2}{2} 
\pm  \lambda^2 v^2 \left|1  +
\left[\alpha  + (1-\alpha)\frac{X\Wm_{\! ,T}}{3\Wm}
\right] \left[\vp + \frac{\sqrt{2} m}{\lambda v^2}\right] \vp
\right|
\nonumber \\  && {}
+ \alpha 
\left(m + \frac{\lambda v^2 \vp}{\sqrt{2}}\right)^2 
+\frac{2 \lambda^2 v^4}{3} \, ,
\label{mpm2}
\eea
For $\alpha \neq 1$ there are additional corrections to the watefall
fields proportional to $\Wm_{\! ,T}$. These are expected to be of the
same size as the other corrections. Hence KKLT-style models are again
mostly ruled out, except for a small range of $\MM$.

We now turn to the behaviour of the moduli fields during inflation. We saw
above how a lower bound on $\MM$ arises from the requirement that 
$V_{\rm max} \gg V_*$. In fact, a stronger bound on $\MM$ comes from
the inflationary corrections to the moduli sector masses. The respective
masses of the real and imaginary parts of $T$, and their fermionic
superpartners are
\be \fl
m_{\Re T}^2 \approx  
\tilde m_T^2 +  \frac{\MM}{m} V_* \, , \qquad 
m_{\Im T}^2 \approx \tilde m_T^2 - \frac{\MM}{m} V_* \, , \qquad 
\tilde m_T^2 \approx \MM^2\left(1+ \frac{\lambda v^2 \vp}{\sqrt{2} m}\right)^2
\label{mT}
\ee
up to $\Or(m)$ corrections. To get the above expressions we have used that
$|\Wm_{\! , T T}|^2 = 3 X \Vm_{{\rm mod}, T \bar T} + \Or(\MM m)$ in
the KL set-up; it should also be remembered that the rescaled coupling
$\lambda$ is modulus dependent.  We have assumed, for simplicity, that $\Wm$
and its derivatives all have the same phase. The masses~\eref{mT} for KKLT
will have different coefficents, but will be qualitatively similar.
Requiring that $\Im T$ is not tachyonic implies either
\be
(a) \quad
\MM \gtrsim \frac{H_*^2}{m}  \qquad \mbox{or} \qquad
(b) \quad \MM \gtrsim \frac{m}{v^2} \gtrsim 2 \times 10^{5} \, m
\label{bound4}
\, .
\ee
For large enough $\MM$, (a) is satisfied by KL- and KKLT-style moduli
sectors, and can in both cases be combined with \eref{bound}. 
The other range (b) is easily satisfied for KL, but not for KKLT.

Finally, we need to check that taking the modulus fixed during inflation, as
assumed above, is a good approximation. As we saw in section~\ref{s:add}, the
modulus dynamics destroys inflation even for the fine-tuned KL set-up when the
modulus and inflation superpotentials are added. For a model with multiplied
superpotentials, this problem is avoided. We will assume that
$\Wm$ and all its dervatives have the same phases. Expanding, much as before,
around the minumum of $\Vm_{\rm mod}$, we take 
$T = T_0 + \delta T_R + \rmi \delta T_I$. Minimising the resulting potential,
we find $\delta T_I =0$ and 
\be
\frac{\delta T_R}{X} \approx - \frac{V_*}{3 m_{\Re T}^2}
 \left(X\frac{D_T \Wm}{\Wm} + 1-\alpha\right)
\ee
giving
\be
-\frac{\delta V_{\rm inf}}{V_*} \approx
\frac{V_*}{3 m^2_{\Re T}}
 \left(\frac{D_T \Wm}{\Wm} +  \frac{1-\alpha}{X}\right)^2 
\lesssim \min\left(\frac{H_*^2}{\MM^2}, \frac{m^2}{\MM^2\vp^2}, 
\frac{m}{\MM} \right)
 \, .
\label{deltaV}
\ee
This is just a small correction to the inflationary potential \eref{VinfM}
provided that either $\MM \gg H_* $, or $\MM \gg m$. At least one of these
conditions is satisfied if we require that  $T$ is not tachyonic during
inflation~\eref{bound4}. 

To summarise, combining the two bounds~\eref{bound} and \eref{bound4} gives
\be
\sqrt{\MM_T m_{3/2}} \gg H_* \gg m_{3/2} \vp_* \, ,
\label{bound2}
\ee
or alternatively
\be
m_{3/2}  \ll \frac{H_*}{\vp_*} \, , \, \MM_T v^2 \, ,
\label{bound3}
\ee
where $\MM_T\approx \MM$ is the mass of $T$ after inflation. Either of
the above bounds can be satisfied by a KL-style scenario without
additional fine-tuning.  KKLT-style models can also satisfy bound
\eref{bound2} and give a viable model of inflation for a limited range
of $\MM$. These conclusions also apply for the more generic,
$\alpha$-dependent K\"ahler~\eref{Ka}.  In both KKLT and KL moduli
stabilisation potentials,  if either of the above bounds is satisfied,
then the modulus does not vary significantly during inflation.  Hence
with only a moderate degree of tuning, inflation can be successfully
combined with a modulus sector when their respective superpotentials
are multiplied.

\section{Inflationary predictions}
\label{s:inf}

Having investigated the effects of the moduli stabilisation sector on the tree
level inflaton potential, we will now determine the moduli corrections to the
one-loop potential. The inflaton slope and curvature, which determine the power
spectrum and the spectral index, are dominated by the one-loop
contribution. This is given explicitly by the Coleman-Weinberg
formula~\eref{CW}. $V_{\rm loop}$ receives contributions from the
non-degenerate boson and fermion pairs, which in our model are not only the
waterfall fields, but also the modulus field $T$ (we will ignore any other
fields for simplicity).  Since the masses are
$\vp$-dependent, their contribution to the loop potential will generate a
non-trivial potential for the inflaton field. In the limit that the slope and
curvature of the inflaton potential is dominated by the waterfall field
contribution to the loop potential, the inflationary predictions are the same
as for the global SUSY model discussed in subsection~\ref{s:hybrid}. We will
then have a working model of inflation. In this section we will determine the
corresponding parameter space. More precise bounds could be obtained by
comparison with the WMAP data, although the results will be sensitive to the
details of the moduli superpotential. Here, we will content ourselves with
order of magnitude bounds. Like the conclusions of the previous
section, our results will apply to the simple K\"ahler~\eref{Kadd},
and to the more generic one~\eref{Ka} for any choice of $\alpha$.

We start by calculating the loop potential.  In the limit that the
gravitino mass is small and the bound \eref{bound} is satisfied, the
expressions for the waterfall masses approach the SUGRA 
results~\eref{mpmSUGRA}. If we further restrict to the regime $\vp < 1$ or
$\lambda \lesssim 0.5$, where the results are manifestly cut-off
independent, we retrieve the global SUSY results \eref{masssplit}.
The loop potential due to the waterfall fields is given by the
familiar expression~\cite{hybrid2}
\be
\fl
V^{(\phi)}_{\rm loop} =\frac{\lambda^2 V_*}{32 \pi^2} 
\[ 2 \ln \( \frac{\lambda^2 v^2 x^2}{\Lambda^2} \)
+(x^2+1)^2 \ln(1+x^{-2}) + (x^2-1)^2 \ln(1-x^{-2}) - 3\]
\label{Vloop}
\ee
with $x^2 = \vp^2/(2v^2)$. Inflation takes place for $x > 1$ and ends as 
$x \to 1$ with the tachyonic instability. 
Using \eref{mT} the modulus contribution to the loop potential is 
\be
\fl
V^{(T)}_{\rm loop} = \frac{V_*^2 \MM^2}{64 \pi^2 m^2} \! 
\[ 2 \ln \! \( \frac{V_* \MM z^2}{\Lambda^2 m} \)
+(z^2+1)^2 \ln(1+z^{-2}) + (z^2-1)^2 \ln(1-z^{-2}) - 3\]
\label{VloopT}
\ee
with 
\be
z^2 = \frac{\tilde m_T^2 m}{V_* \MM} = 
\frac{\MM}{m}\left(\frac{m}{\lambda v^2}+ \frac{\vp}{\sqrt{2}}\right)^2 \, .
\ee  

The loop potential gives a negligible contribution to the total energy density
during inflation $V_*$, but it is the dominant contribution to the slow-roll
parameters $\epsilon$ and $\eta$.  Hence to see whether it is the waterfall or
the modulus contribution to the potential which dominates the inflationary
dynamics, we have to compare their derivatives. In addition we need to satisfy
the upper bound on $m$~\eref{bound}, so that neglecting $\Or(m)$ terms is a
good approximation. Requiring that the axion is non-tachyonic during
inflation gives a further, lower bound on the modulus mass scale
$\MM$~\eref{bound4}. Finally, we note that both KKLT and KL moduli
stabilisation potentials have  $m \lesssim \MM$, which restricts the
allowed parameter space. If the above constraints are satisfied, then
the modulus automatically remains fixed during inflation, and its
dynamics do not produce further constraints.

We expect to retrieve standard hybrid inflation results in the limit that the
mass splitting between the modulus field and its superpartners is small,  as
this sets the overall scale of the modulus loop potential.  In this limit 
$z^2 \gg 1$. The $\vp$-dependence only
enters $V^{(T)}_{\rm loop}$ via $\tilde m_T^2$, and we find it convenient to
write 
\be
\tilde m_T = \MM(1+\delta_m) \, ,  \quad {\rm with} \quad
\delta_m = \frac{\lambda v^2 \vp}{\sqrt{2} m} \, .
\label{dm}
\ee
The modulus loop effects are 
suppressed in the limit $\delta_m \to 0$.  As it turns out the 
$\delta_m \to 0$ limit can be relaxed, and it will be sufficient to consider
the loop potential in the regime $z^2 \gg 1$ in order to determine the allowed
parameter space. The modulus contribution in the large $z$-limit is
\be
\left(V_{\rm loop}^{(T)}\right)'_* \approx  
\frac{\lambda^5 v^{10}\MM^2}{16\sqrt{2} \pi^2 m^3} \frac1{(1+\delta_m)} \, .
\label{dVT}
\ee
This is to be compared with the equivalent expression for the waterfall field
potential. 

\subsection{Large coupling, $\lambda^2 \gtrsim 10^{-5}$}

In the large coupling regime, $\lambda^2 > 7.4 \times 10^{-6}$, we can
approximate \eref{Vloop} by the large $x$ result \eref{Vlarge} and 
\be
\lim_{x\gg 1}
\left(V_{\rm loop}^{(\phi)}\right)'_* \approx 
\frac{\lambda^3 v^4}{4\pi\sqrt{N_*}} \, ,
\ee
where we used $\vp_* \approx \lambda \sqrt{N_*}/(2\pi)$.  This dominates over
\eref{dVT} for 
\be
\MM^2 < \frac{4\sqrt{2} \pi m^3 (1+\delta_m)}{\sqrt{N_*} \lambda^2 v^6}
\approx\left\{
\begin{array}{ll}
1.3 \times 10^{16} \, {m^3} {\lambda^{-2}} \, ,  &\delta_m \ll 1 \\
6.4 \times 10^{10} \, m^2 \, ,         &\delta_m \gg 1 
\end{array}
\right.
\label{b1}
\ee
where we used $v^2 \approx 6 \times 10^{-6}$ and $N_* = 60$.  Small 
$m < 4.8 \times 10^{-6} \lambda^2$ corresponds to the large $\delta_m > 1$ 
regime. This should be combined with the axion mass bound~\eref{bound4} which
translates to
\be
\MM > \frac{\lambda^2 v^4}{m(1+\delta_m)^2}
\approx\left\{
\begin{array}{ll}
3.1 \times 10^{-11} \, m^{-1} \lambda^{2}\, ,  &\delta_m \ll 1 \\
1.3  \, m \lambda^{-2} \, ,         &\delta_m \gg 1 
\end{array}
\right.
\ee
and the bound from modulus corrections to $m_\pm$~\eref{bound}, which
gives $m < 1.4 \times 10^{-5}$.

\begin{figure}
\centerline{ 
\includegraphics[width=8cm]{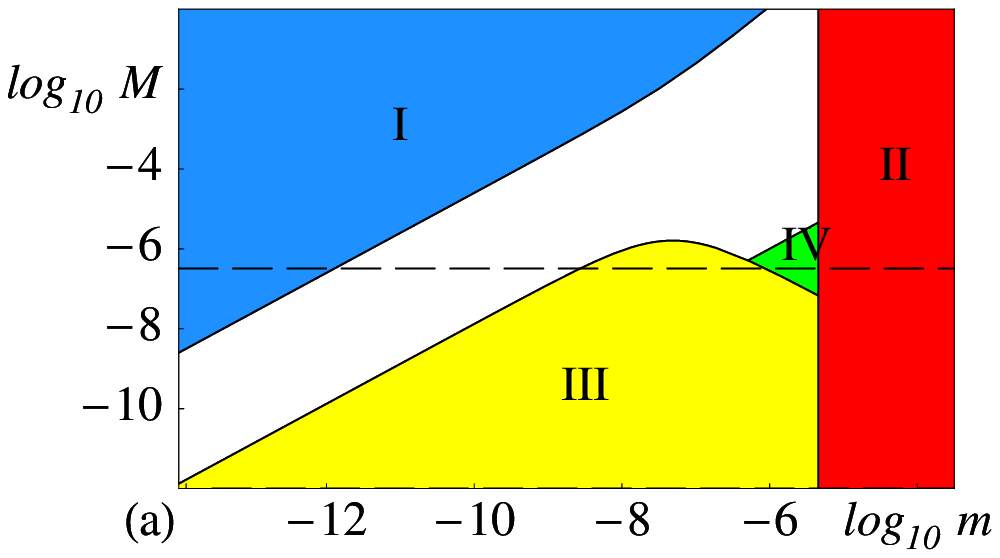} 
\includegraphics[width=8cm]{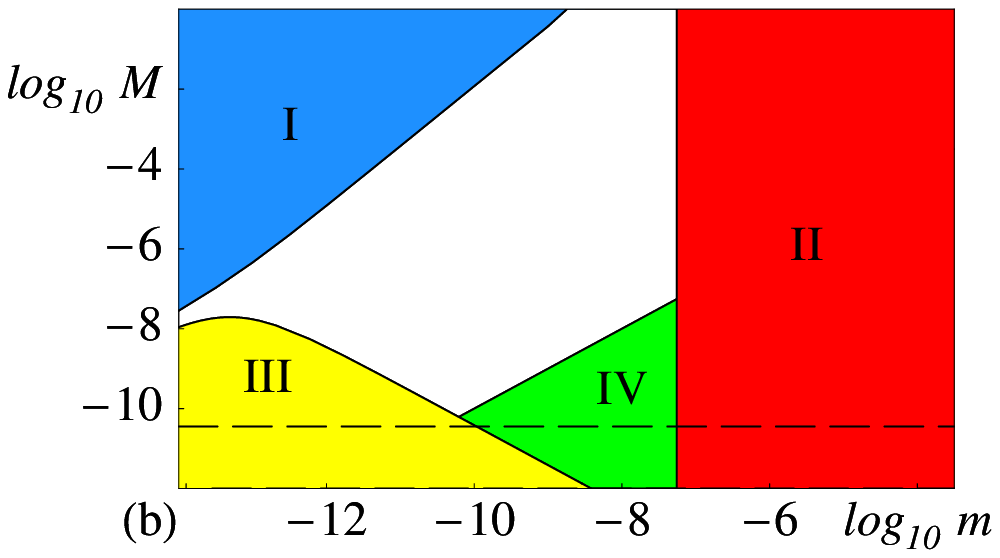}
}
\caption{Parameter space in $\{ \log_{10}(m), \log_{10}(\MM)\}$ for
  (a) $\lambda = 0.1$ and (b) $\lambda = 10^{-4}$. In the
  white region the model reduces to SUSY hybrid inflation.  Regions I-IV are
  excluded, because I: the modulus mass dominates the 1-loop potential, II: the
  gravitino mass is too large, III: the modulus is tachyonic during inflation, 
  and IV: the modulus potential property $m \lesssim \MM$ is not satisfied. 
  The dashed lines correspond to $H_*=\MM$}
\label{F:l1}
\end{figure}

The parameter space in the $\{\log_{10}(m), \log_{10}(\MM)\}$-plane is shown
for $\lambda = 0.1$ in figure~\ref{F:l1}a.  In the white region the
inflationary results approach those of the global SUSY model discussed in
section~\ref{s:hybrid}.  Hence, there is a region of parameter space for which
multiplying superpotentials gives a viable model of $F$-term hybrid inflation.
This is in sharp contrast to a combined model in which the superpotentials are
summed: as we saw in section~\ref{s:add}, inflation fails in this case.

In all of parameter space $z^2 \gg 1$, and our analytic results are
valid.  In region I the loop potential is dominated by the modulus
contribution~\eref{b1}; when this becomes too large inflation is
ruined.   In region II the bound~\eref{bound} on the gravitino mass is
violated, and moduli corrections are too large for successful
inflation.  Region III is excluded as it gives a tachyonic
axion~\eref{bound4}. Except for very near the border with region III
the $\eta$-parameter is dominated by the waterfall field  contribution
to the loop potential. Finally, region IV bounds $m \lesssim \MM$
which is a property of both KKLT and KL-style moduli sectors.  Viable,
KKLT-style models correspond to the upper-left edge of region
IV. Since this class of models has only one mass scale $\MM\sim m$,
it corresponds to a line in the plotted, two-dimensional parameter
space. The fact that $\vp_* < 1$ during inflation allows \eref{bound}
and \eref{bound4} to be realised simultaneously for a limited range of
$\MM$ (which increases in size as coupling $\lambda$ is reduced). The
two-scale KL model works throughout the white region of parameter
space in the plot.

In the $\delta_m \gg 1$ regime the effective modulus mass is enhanced during
inflation compared to its vacuum value $\MM$, as can be seen 
from~\eref{enhanced}. This allows for the possibility of having
$m < \MM < H_*$, yet with the modulus fixed during inflation. For
$\lambda = 0.1$ the inflationary scale is $H_* \approx 10^{-6}$.  The dashed
lines in figure~\ref{F:l1} correspond to $\MM = H_*$; we see that indeed 
$\MM < H_*$ is realised in large part of parameter space, contrary to
naive expectations.

\subsection{Small coupling, $\lambda^2 \lesssim 10^{-5}$}

We can apply the same analysis for the small coupling regime 
$\lambda^2 < 7.4 \times 10^{-6}$.  In this case $\vp_* \approx \sqrt{2} v$ and 
$v^2 = 5.6 \times 10^{-6} [\lambda^2/(7.4 \times 10^{-6})]^{1/3}$.  In the
small $x \to 1$ limit the slope of the waterfall loop potential becomes
\be
\lim_{x \to 1}
\left(V_{\rm loop}^{(\phi)}\right)' 
\approx \frac{\lambda^4 v^3 \log(2)}{4\sqrt{2} \pi^2}
\ee
which is to be compared with \eref{dVT}. The waterfall field contribution
dominates the one-loop potential for 
\be
\MM^2 < \frac{4\log(2)(1+\delta_m) m^3}{\lambda v^7}
\approx\left\{
\begin{array}{ll}
6.9 \times 10^{12} \, {m^3} {\lambda^{-10/3}},  &\delta_m \ll 1 \\
3.4 \times 10^{7} \, m^2 \lambda^{-4/3} ,         &\delta_m \gg 1 
\end{array}
\right.
\ee
Small $m < 4.9 \times 10^{-6} \lambda^2$ corresponds to the large
$\delta_m > 1$ regime. This has to be combined with
$m < 1.2 \times 10^{-2} \lambda^{4/3}$ from \eref{bound} and
\be
\MM > \left\{
\begin{array}{ll}
8.3 \times 10^{-8} \, m^{-1} \lambda^{10/3} ,  &\delta_m \ll 1 \\
3.4 \times 10^{3} \, m \lambda^{-2/3} ,         &\delta_m \gg 1 
\end{array}
\right.
\ee
from \eref{bound4}.  The results for $\lambda = 10^{-4}$ are shown in
figure~\ref{F:l1}b. We see that for smaller couplings the modulus
stabilisation scale needs to be larger than the Hubble scale
during inflation.  E.g.\ for  $\lambda = 10^{-4}$ the inflationary scale is 
$H_* \approx 10^{-10}$, and  $\MM > H_*$ in all of parameter space for
successful inflation. This contrasts with the situation for larger couplings,
as we saw in the previous subsection.

\section{Conclusions}

The flatness of the inflationary potential in SUGRA models is
typically spoilt by corrections coming from supersymmetry
breaking. Ironically enough, the vacuum energy which drives inflation 
breaks SUSY spontaneously, and so gives soft corrections to the inflaton;
this is the well-known $\eta$-problem. Introducing a shift symmetry for
the inflaton will protect the inflation sector from itself, and remove
the problem. However there will still be corrections coming from other
sectors of the full theory, which can also disrupt inflation. In this paper we
studied the effects of a moduli stabilisation sector on a $F$-term
SUGRA hybrid inflation model.  

We considered both a KKLT-like moduli stabilisation scheme, in which there is
only one scale in the potential so $m_T \sim m_{3/2}$, as well as a
fine-tuned two-scale KL-like set-up with $m_T \gg m_{3/2}$. In the KKLT
set-up, requiring the modulus to be fixed during inflation raises the scale of
the modulus potential, and as a result the soft corrections to both the
inflaton slope and the waterfall field masses are too large for inflation to
take place.  This problem is circumvented in the KL set-up where the gravitino
mass, and consequently the corrections to the inflationary potential, can be
tuned arbitrarily small.

One would be inclined to conclude that KL moduli stabilisation can be combined
almost effortlessly with inflation.  But this is not true.  The above
conclusions only hold in the limit that the modulus field remains fixed during
inflation.  Although this seems like a good approximation, as the
displacement of the modulus minimum during inflation is indeed small, the
correction to the flat inflaton potential is nevertheless large.  In fact, it
gives $\eta \approx -3$, and thus no slow-roll inflation.  This analysis shows
that it is important to take the dynamics of all fields during inflation into
account, otherwise crucial effects may be missed.

We have proposed a way to solve all of the above problems, and
successfully combine $F$-term hybrid inflation with moduli
stabilisation.  The idea is to combine the modulus and inflaton
sectors not by adding their respective superpotentials, as is usually
done, but by adding their respective K\"ahler functions $G = K + \ln|W|^2$ 
instead.  Adding  K\"ahler functions corresponds to adding
K\"ahler potentials and multiplying superpotentials.  This way of
combining sectors greatly reduces their interactions.  In particular,
for the case of combining inflation with a modulus sector, it greatly 
reduces the displacement of the modulus during inflation.
Consequently the correction to the inflationary potential is
harmlessly small.  For the fine-tuned two-scale KL set-up, or for a
one-scale KKLT set-up with a fine-tuned mass scale, the corrections to
the inflaton slope and waterfall masses are small as well.  Hence we indeed
succeeded in constructing a successful model of inflation in the
presence of moduli.

Even when multiplying superpotentials, there are still some constraints on the
moduli sector parameters for viable inflation.  The graviton mass should be
small enough to suppress the moduli corrections during inflation.  The modulus
mass needs to be heavy and non-tachyonic during inflation to remain
stabilised.  Finally the loop potential should be dominated by the
contribution of the waterfall fields rather than by the modulus contribution.
Nevertheless, there is still a large region of gravitino and modulus mass
scales for which inflation works, and the inflationary predictions are nearly
indistinguishable from the global SUSY model in the absence of moduli fields.

\ack
We are both grateful to K. Sousa and particularly A. Ach\'ucarro for useful
discussions, inspiration and for the suggestion that multiplication of
superpotentials could be natural and helpful. We also thank N. Bevis,
C. Burgess and J. Rocher for a useful comments.  SCD thanks the Netherlands Organisiation for Scientific Research (NWO) for financial support.


\section*{References}

\begin{thebibliography}{19}
\bibitem{anaXW}
 A.~Ach\'ucarro and K.~Sousa,
  {\em F-term uplifting and moduli stabilization consistent with Kahler
  invariance,}
  0712.3460 [hep-th]
  %%CITATION = ARXIV:0712.3460;%%

\bibitem{copeland}
  E.~J.~Copeland, A.~R.~Liddle, D.~H.~Lyth, E.~D.~Stewart and D.~Wands,
  {\em False vacuum inflation with Einstein gravity,}
  Phys.\ Rev.\  D {\bf 49} (1994) 6410
  [astro-ph/9401011]

\bibitem{dine}
  M.~Dine, L.~Randall and S.~D.~Thomas,
  {\em Supersymmetry breaking in the early universe,}
  Phys.\ Rev.\ Lett.\  {\bf 75} (1995) 398
  [hep-ph/9503303]

\bibitem{KKLT}
S.~Kachru, R.~Kallosh, A.~Linde and S.~P.~Trivedi,
  {\em De Sitter vacua in string theory,}
  Phys.\ Rev.\  D {\bf 68}, 046005 (2003)
  [hep-th/0301240]

\bibitem{KL2}
R.~Kallosh and A.~Linde,
  {\em Landscape, the scale of SUSY breaking, and inflation,}
  JHEP {\bf 0412} (2004) 004
  [hep-th/0411011]

\bibitem{modular}
T.~Banks, M.~Berkooz, S.~H.~Shenker, G.~W.~Moore and P.~J.~Steinhardt,
  {\em Modular Cosmology,}
  Phys.\ Rev.\  D {\bf 52} (1995) 3548
  [hep-th/9503114]

\bibitem{rt}
  J.~J.~Blanco-Pillado {\it et al.},
  {\em Racetrack inflation,}
  JHEP {\bf 0411} (2004) 063
  [hep-th/0406230]
\nonum
J.~J.~Blanco-Pillado {\it et al.},
  {\em Inflating in a better racetrack,}
  JHEP {\bf 0609} (2006) 002
  [hep-th/0603129]
\nonum
 J.~P.~Conlon and F.~Quevedo,
  {\em Kaehler moduli inflation,}
  JHEP {\bf 0601} (2006) 146
  [hep-th/0509012]
\nonum
P.~Brax, A.~C.~Davis, S.~C.~Davis, R.~Jeannerot and M.~Postma,
  {\em D-term Uplifted Racetrack Inflation,}
  0710.4876 [hep-th]
%%CITATION = ARXIV:0710.4876;%%


\bibitem{KKLMMT}
S.~Kachru, R.~Kallosh, A.~Linde, J.~M.~Maldacena, L.~P.~McAllister and
S.~P.~Trivedi, 
  {\em Towards inflation in string theory,}
  JCAP {\bf 0310} (2003) 013
  [hep-th/0308055]

\bibitem{baumann}
D.~Baumann, A.~Dymarsky, I.~R.~Klebanov and L.~McAllister,
  {\em Towards an Explicit Model of D-brane Inflation,}
  0706.0360 [hep-th].
  %%CITATION = ARXIV:0706.0360;%%

\bibitem{krause}
  A.~Krause and E.~Pajer,
  %``Chasing Brane Inflation in String-Theory,''
  arXiv:0705.4682 [hep-th].
  %%CITATION = ARXIV:0705.4682;%%
  
\bibitem{gaillard}
  M.~K.~Gaillard, D.~H.~Lyth and H.~Murayama,
  {\em Inflation and flat directions in modular invariant superstring  
  effective theories,}
  Phys.\ Rev.\  D {\bf 58} (1998) 123505
  [hep-th/9806157]

\bibitem{shift2}
 J.~P.~Hsu, R.~Kallosh and S.~Prokushkin,
  {\em On brane inflation with volume stabilization,}
  JCAP {\bf 0312} (2003) 009
  [hep-th/0311077]

\bibitem{KL}
R.~Kallosh and A.~Linde,
  {\em O'KKLT,}
  JHEP {\bf 0702} (2007) 002  [hep-th/0611183]

\bibitem{mfi}
P.~Brax, C.~van de Bruck, A.~C.~Davis and S.~C.~Davis,
  {\em Coupling hybrid inflation to moduli,}
  JCAP {\bf 0609} (2006) 012
  [hep-th/0606140]

\bibitem{lalak}
 Z.~Lalak and K.~Turzynski,
  {\em Back-door fine-tuning in supersymmetric low scale inflation,}
  0710.0613 [hep-th]
  %%CITATION = ARXIV:0710.0613;%%

\bibitem{mdi}
Ph.~Brax, C.~van de Bruck, A.~C.~Davis, S.~C.~Davis, 
R.~Jeannerot and M.~Postma,
  {\em Moduli corrections to D-term inflation,}
  JCAP {\bf 0701} (2007) 026  [hep-th/0610195]

\bibitem{riotto}
  N.~Bartolo and A.~Riotto,
  {\em Possibly Large Corrections to the Inflationary Observables,}
  0711.4003 [astro-ph]
  %%CITATION = ARXIV:0711.4003;%%

\bibitem{binXW}
P.~Binetruy, G.~Dvali, R.~Kallosh and A.~Van Proeyen,
  {\em Fayet-Iliopoulos terms in supergravity and cosmology,}
  Class.\ Quant.\ Grav.\  {\bf 21} (2004) 3137
  [hep-th/0402046]

\bibitem{hybrid1}
  A.~D.~Linde,
  {\em Hybrid inflation,}
  Phys.\ Rev.\  D {\bf 49} (1994) 748  [astro-ph/9307002]

\bibitem{hybrid2}
  G.~R.~Dvali, Q.~Shafi and R.~K.~Schaefer,
  {\em Large scale structure and supersymmetric inflation without fine tuning,}
  Phys.\ Rev.\ Lett.\  {\bf 73} (1994) 1886  [hep-ph/9406319]

\bibitem{CW}
  S.~R.~Coleman and E.~Weinberg,
  {\em Radiative Corrections As The Origin Of Spontaneous Symmetry Breaking,}
  Phys.\ Rev.\  D {\bf 7} (1973) 1888

\bibitem{loop}
  S.~Ferrara, C.~Kounnas and F.~Zwirner,
  {\em Mass formulae and natural hierarchy in string effective supergravities,}
  Nucl.\ Phys.\ B {\bf 429} (1994) 589
  [Erratum-ibid.\ B {\bf 433} (1995) 255]  [hep-th/9405188]

\bibitem{kibble}
 T.~W.~B.~Kibble,
  {\em Topology Of Cosmic Domains And Strings,}
  J.\ Phys.\ A  {\bf 9} (1976) 1387

\bibitem{strings}
R.~Jeannerot, J.~Rocher and M.~Sakellariadou,
  {\em How generic is cosmic string formation in SUSY GUTs,}
  Phys.\ Rev.\  D {\bf 68} (2003) 103514
  [hep-ph/0308134]

\bibitem{COBE}
E.~F.~Bunn, A.~R.~Liddle and M.~J.~White,
  {\em Four-year COBE normalization of inflationary cosmologies,}
  Phys.\ Rev.\  D {\bf 54} (1996) 5917
  [astro-ph/9607038]

\bibitem{WMAP3} 
D.~N.~Spergel {\it et al.}  [WMAP Collaboration],
  {\em Wilkinson Microwave Anisotropy Probe (WMAP) three year results:
  Implications for cosmology,}
  Astrophys.\ J.\ Suppl.\  {\bf 170} (2007) 377  [astro-ph/0603449]

\bibitem{urrestilla}
  N.~Bevis, M.~Hindmarsh, M.~Kunz and J.~Urrestilla,
   {\em Fitting Cosmic Microwave Background Data with Cosmic Strings and
Inflation,}
   Phys.\ Rev.\ Lett. {\bf 100} (2008) 021301
   [astro-ph/0702223]

\bibitem{wyman}
  M.~Wyman, L.~Pogosian and I.~Wasserman,
  {\em Bounds on cosmic strings from WMAP and SDSS,}
  Phys.\ Rev.\  D {\bf 72}, 023513 (2005)
  [Erratum-ibid.\  D {\bf 73}, 089905 (2006)]
  [astro-ph/0503364]

\bibitem{bjorn}
  R.~A.~Battye, B.~Garbrecht and A.~Moss,
  {\em Constraints on supersymmetric models of hybrid inflation,}
  JCAP {\bf 0609}, 007 (2006)
  [astro-ph/0607339]

\bibitem{mairi}
  J.~Rocher and M.~Sakellariadou,
  {\em Supersymmetric grand unified theories and cosmology,}
  JCAP {\bf 0503} (2005) 004
  [hep-ph/0406120]

\bibitem{rachel}
  R.~Jeannerot and M.~Postma,
  {\em Confronting hybrid inflation in supergravity with CMB data,}
  JHEP {\bf 0505}, 071 (2005)
  [hep-ph/0503146]

\bibitem{semilocal}
  J.~Urrestilla, A.~Ach\'ucarro and A.~C.~Davis,
  {\em D-term inflation without cosmic strings,}
  Phys.\ Rev.\ Lett.\  {\bf 92} (2004) 251302
  [hep-th/0402032]

\bibitem{hybridsugra}
  A.~D.~Linde and A.~Riotto,
  {\em Hybrid inflation in supergravity,}
  Phys.\ Rev.\  D {\bf 56} (1997) 1841  [hep-ph/9703209]

\bibitem{gkp}
  S.~B.~Giddings, S.~Kachru and J.~Polchinski,
  {\em Hierarchies from fluxes in string compactifications,}
  Phys.\ Rev.\  D {\bf 66} (2002) 106006  [hep-th/0105097]

\bibitem{bqk}
 C.~P.~Burgess, R.~Kallosh and F.~Quevedo,
  {\em de Sitter string vacua from supersymmetric D-terms,}
  JHEP {\bf 0310} (2003) 056  [hep-th/0309187]

\bibitem{ana}
 A.~Ach\'ucarro, B.~de Carlos, J.~A.~Casas and L.~Doplicher, {\em de Sitter
 vacua from uplifting D-terms in effective supergravities from
 realistic strings,} JHEP {\bf 0606} (2006) 014
 [hep-th/0601190]

\bibitem{O'Raifeartaigh}
  L.~O'Raifeartaigh,
  {\em Spontaneous Symmetry Breaking For Chiral Scalar Superfields,}
  Nucl.\ Phys.\  B {\bf 96} (1975) 331

\bibitem{ISS}
  K.~Intriligator, N.~Seiberg and D.~Shih,
   {\em Dynamical SUSY breaking in meta-stable vacua,}
  JHEP {\bf 0604} (2006) 021  [hep-th/0602239]

\end{thebibliography}
\end{document}